\documentclass[5p,times]{elsarticle}

\usepackage{lineno}
\usepackage{hyperref}
\usepackage{svg}
\usepackage{lineno}
\usepackage{xcolor}

\journal{Automation in Construction}

\begin{document}

\begin{frontmatter}

\title{BIM-to-BRICK: Using graph modeling for IoT/BMS and spatial semantic data interoperability within digital data models of buildings}
\author{Filippo Vittori \textsuperscript{a,b}, Chuan Fu Tan \textsuperscript{d}, Anna Laura Pisello* \textsuperscript{a,b}, \\Adrian Chong \textsuperscript{c}, Clayton Miller* \textsuperscript{c}}

\address{\textsuperscript{a}Department of Engineering, University of Perugia, Perugia, PG, Italy\\\textsuperscript{b}CIRIAF – Interuniversity Research Centre, University of Perugia, Perugia, PG, Italy\\\textsuperscript{c}College of Design and Engineering, National University of Singapore, Singapore\\\textsuperscript{d}Open Blue Innovation Centre, Johnson Controls International, Singapore\\
$^*$Corresponding Authors:\\ clayton@nus.edu.sg, +65 81602452\\anna.pisello@unipg.it, +39 339 692 7839}

\begin{abstract}
The holistic management of a building requires data from heterogeneous sources such as building management systems (BMS), Internet-of-Things (IoT) sensor networks, and building information models. Data interoperability is a key component to eliminate silos of information, and using semantic web technologies like the BRICK schema, an effort to standardize semantic descriptions of the physical, logical, and virtual assets in buildings and the relationships between them, is a suitable approach. However, current data integration processes can involve significant manual interventions. This paper presents a methodology to automatically collect, assemble, and integrate information from a building information model to a knowledge graph. The resulting application, called BIM-to-BRICK, is run on the SDE4 building located in Singapore. BIM-to-BRICK generated a bidirectional link between a BIM model of 932 instances and experimental data collected for 17 subjects into 458 BRICK objects and 1219 relationships in 17 seconds. The automation of this approach can be compared to traditional manual mapping of data types. This scientific innovation incentivizes the convergence of disparate data types and structures in built-environment applications.

\end{abstract}

\begin{keyword}
Digital twin \sep Building Management System \sep Automation \sep Occupants’ behavior \sep Data interoperability \sep Building operation
\end{keyword}
\end{frontmatter}

% \linenumbers
% \modulolinenumbers[5]

\section{Introduction}
\label{sec:intro}
Building management entails operating various systems and governing different data sources about operations, monitoring, maintenance, energy management, auditing, occupant comfort, safety, and security. Central to these processes is the Building Management System (BMS), an overarching computer-based control system that relays information from different building subsystems. It provides data from the core operational systems of a building to monitor, configure and regulate those subsystems. Internet of Things (IoT) devices can complement and support these efforts as portable sensors allow for ancillary functions and granular monitoring of facilities operations. Recently, Human-as-a-Sensor (HaaS), serving as a supplementary data source, has shown effectiveness in evaluating and optimizing the human comfort of the building~\cite{Jayathissa2020-pv}. These can be in the form of digital surveys on wearable platforms that can provide direct and more accurate feedback on various environmental, physiological, and psychological parameters ~\cite{Jayathissa2019, ABDELRAHMAN2022109090}. Furthermore, Building Information Modeling (BIM) can supplement the as-mentioned data with geometrical, spatial, and contextual information extracted from a 3D model. 

The aggregation of these data sources in a collaborative process can be expressed as a digital twin, a concept that was brought to light for the first time in 2015 ~\cite{ROSEN2015567} when NASA’s engineers and scientists created two identical capsules. One of the capsules was sent to outer space, while the twin was kept on earth to simulate the behavior and monitor the situation of the sent apparatus. Currently, there is no standard definition of the \emph{digital Twin} concept as it depends on the context~\cite{Wagner2019}. The various existing definitions are related to specific characteristics derived from multiple use cases. The understanding of \emph{digital twins} has continuously changed during its development ~\cite{van2020taxonomy}. Although the definition of the concept is still under development, the topic is gaining momentum and rising interest as one of the multiple forms of the current digital transformation. Digital Twins are used in several fields. Already in 2015, Rosen et al. \cite{ROSEN2015567} investigated the inherence of the topic within the manufacturing field, focusing attention on Digital Twins’ potential within the context of Industry 4.0 and CPS (cyber-physical system) modularity. This kind of approach is probably the most simple and immediate, but it represents only the baseline within the framework. Intelligent transport systems represent other methodology applications, providing an example of a different conception of targeted entities: not only objects can be mirrored and become part of a Digital Twin. Rudskoya et al. ~\cite{rudskoy2021digital} introduced the concept within Intelligent Transport Systems to solve the main problems of the transport network and effectively develop it. In this context, the purpose of the digital twin adopted emerges from the building industry’s demands to monitor and control assets closely, providing real-time insights into a building’s performance and events ~\cite{Boje2020}. It increases the dimensionality of information by being able to relate to data in a representative manner visually and better improve the operations and performance of various activities in the building’s lifecycle ~\cite{buildings12020120, MEONI2022104220, Miller_2021}. In achieving such an integration, better interoperability of those heterogeneous data sources is necessary but often lacking in existing buildings. Information on point naming conventions in BMS, drawings, IoT devices, and other details are typically not standardized and hard to interpret without first-hand knowledge ~\cite{SATTLER2019271,enshassi2019limitation}. This situation has also led to other issues, such as becoming a significant barrier to the deployment of building applications for automated commissioning, fault detection, and diagnostic and optimization of operations ~\cite{10.1145/2821650.2821674, su14105810}. 

Often, a manual and time-intensive process of mapping and interpreting the necessary data for these software tools has to be carried out ~\cite{10.1145/2821650.2821674, su13073930, 10.1145/3276774.3276795}. This has led to hesitancy, delays, and additional costs~\cite{en14082090}~\cite{Muller2015/04} in adopting more innovative, energy-efficient, and optimized building operations, which can mean significant undesired consequences by the infrastructure landscape on a national or global level ~\cite{su13084496,gcr2004cost}. Easing the process of adopting national and international standards that define entities and concepts in interoperable, machine-readable semantic models for buildings is thus critical to minimize this issue. Doing so can also facilitate and encourage the adoption of more intelligent, modern technologies, increasing cost-savings for building management in the long run,~\cite{su13094658}. As such, this project aims to create a convenient process for buildings to onboard one of those standards that enable data interoperability and serve as a common platform for the buildings’ metadata. A plugin in Revit has been created to convert BIM to BRICK schema, automating the aggregation and mapping of all the available information. It has been successfully demonstrated on SDE4, a National University of Singapore building, combining its architectural, mechanical systems, human comfort, and BMS information.

\subsection{Data integration in buildings}
Moving towards more holistic and innovative management of buildings requires integrating data from diverse sources relating to their operations and space. It presents a need to eliminate silos of building information, such as from BIM, mechanical systems, floor plans, IoT devices, BMS, and occupants, to create a single pane of glass perspective of its operations and lifecycle processes. As such, data interoperability is an essential enabler to this effort by harmonizing contextual information of the different data domains and bringing them to a shared environment. This unification can allow a virtual representation of the building in a digital twin representation. In this regard, semantic web technologies like ontologies and resource description frameworks (RDF) can facilitate the standardization of entities, concepts, and relationships to realize a shared understanding of knowledge in a particular domain~\cite{STUDER1998161,suwanmanee2005owl}. The semantic information and standardized syntax can be embodied in knowledge graphs, represented in their nodes and edges. Its usage provides an avenue for creating Linked Data~\cite{isaac2011library} through integrating and extending to other heterogeneous information and ontologies. 

Given its potential in the built environment, several building-related ontologies using semantic web technologies~\cite{en14072024} have been introduced in the recent decade, such as Project Haystack~\cite{john2020project}, Building Topology Ontology (BOT) ~\cite{rasmussen2021bot}, RealEstateCore ~\cite{hammar2019realestatecore}, Digital Twin Language ~\cite{gupta2021hierarchical}, Smart Appliances Reference Ontology ~\cite{daniele1smart}, and BRICK schema ~\cite{schema2016brick}. Within this work, the BRICK schema is chosen because it includes a tagging system similar to Project Haystack that augments tags with formal semantic rules that promote consistency and interoperability, it can be easily integrated with other ontologies (i.e. Building Topology Ontology and Smart Appliances Reference Ontology), and it is enriched with dedicated classes to represent occupants. In addition to the cited reasons, BRICK enhances the IFC capabilities of relationship description. Brick builds upon prior work and introduces a number of novel concepts. Brick uses easy-to-understand Tags and TagSets to specify sensors and subsystems in a building ~\cite{balaji2018brick}. Garrido-Hidalgo et al. provide a solid example of the integration of BRICK with other models, interlinking the overall BRICK data model to five other ontologies. Further noticeable integration involving the BIM framework is carried out by Zhu et al., who focus the research on information models for highway operational risk management based on the IFC-Brick schema ~\cite{zhu2023information}.

\subsection{Current approaches to data integration}
Before syntax and metadata of building operations can be expressed in one of these formal ontologies and specifications, a transitional process from their existing representations is required. Formulating semantic graphs for building operations is often challenging in having to infer from existing metadata that often lack consistency or are based on arbitrary naming conventions~\cite{BALAJI20181273, 10.1145/2821650.2821669}. For instance, building management systems (BMS) point labels are a primary source of building operations metadata. Even though there are existing open-source tools such as OpenRefine~\cite{verborgh2013using}, a significant manual effort to map them to RDF representation can still be required due to several irreconcilable definitions~\cite{WANG2018107}. The manual inference process requires understanding the context of each entity in the building and relating it to its associated sub-system and spatial information. This can be further exacerbated by including data from standalone IoT devices or disparate sources such as human comfort parameters from HaaS. Current efforts in unifying operational and spatial information in some of these ontologies often involve an additional process of manual cross-linking, mapping, or inserting a common identifier of the different datasets~\cite{WANG2018107, Chamari_Petrova_Pauwels_2022, farghaly2019bim, su13073930}. For instance, Chamari et al. developed an approach of integrating Industry Foundation Classes (IFC) and BMS as RDF~\cite{Chamari_Petrova_Pauwels_2022}. However, a mapping table containing all the entities and a text file defining the relationships has to be created to facilitate the process. 

Farghaly et al. presented a method for integrating building-related assets, Linked Data, and BIM. To cross-link the different data sets, the authors manually map the other assets to their related information. The reliance on human reconciliation of these buildings' assets is further highlighted by Weimin et al., who pointed out that existing point names and other point attributes of buildings were not easily decodable by machines~\cite{WANG2018107}. In their efforts to combine data from Sustainment Management System with Revit, Loeh et al. have to manually enter a common identifier for each component of BIM ~\cite{su13137014}. Manual rework is also necessary for Dave et al. to integrate IFC with IoT devices through Open Messaging interfaces~\cite{DAVE201835}. Without standardized export guidelines for the web, it is tedious to reconcile objects in IFC with sensor data. These manual interventions can add additional complexity and be susceptible to errors in how they are implemented.

\subsection{Objective and scope}
To provide a more intuitive and automated conception of a standard metadata environment, this work introduces a process to generate an RDF graph of all available spatial, geometrical, mechanical, occupants, and operational information in the BIM of a building in an automated manner. It eliminates the need for the extensive and manual transitional and multi-layered integrative process employed with current methodologies. The result of this work is a convenient plugin created in Revit using Dynamo to automatically infer and extract all the included entities, their relationships, and data availability and subsequently construct an integrated knowledge graph. Moreover, with the inclusion of the BIM authoring software identifiers, a bidirectional link between Revit and BRICK is included in the generated model. This helps in the association of BRICK data back to Revit for relaying any changes or updates to building operations.
The requirements to be addressed by the developed plug-in regarding information collection involve the objects created directly both inside the targeted authoring BIM platform and gathered from external sources and associated relationships. A crucial requirement for the methodology is the capability to access a targeted ontology directly inside the environment to translate the data structure deriving from the aforementioned sources. The main requirement for the developed software is to eliminate the manual work of converting the information collected into the targeted ontology-based resulting model and generate an ontology-based output to be visualized in a dedicated viewer. Furthermore, the ability to update the generated ontology-based model by simply re-running the application and to automatically include in the generated ontology-based output relevant information and parameter values to unequivocally relate every single object to the BIM version of it represent additional features included to streamline the process and improve the obtained framework proposed.

\begin{figure}[h]
\begin{center}
\includegraphics[width= 0.45\textwidth, trim= 0cm 0cm 0cm 0cm,clip]{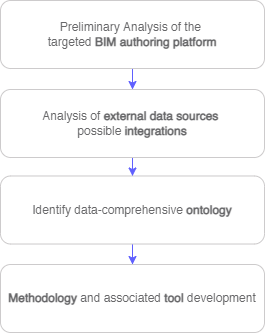}
\caption{Main tasks performed to accomplish the final result of the current work}
\label{fig:TaskWorkflow}
\end{center}
\end{figure}

In summary, the final objective is to streamline the process of linking metadata generated in BIM to RDF-based data comprehensive environments by automating the generation of an RDF graph (BRICK-based) and constructing an integrated knowledge graph. Furthermore, the objective aims to establish a bidirectional link between the BIM source and the BRICK model, to enable efficient communication and updates between the two frameworks. To accomplish this objective, the first task performed within this work is a preliminary analysis of the modeling capability of the targeted BIM authoring platform, both in terms of punctual objects and the relationship between objects (i.e. logical relationship between elements, system modeling capability, the spatial distribution of objects, etc.). The second step performed involved the analysis of possible ways to integrate additional sources and external sources of data inside the authoring platform. The third step of the work was focused on the analysis of possible ontologies to standardize and embrace in a comprehensive environment the data treated within the BIM (and external sources) environment. The fourth step performed consisted of a comparative analysis of the current possible ways of translating and linking the data between the targeted BRICK ontology and the BIM environment, with manual conversion and no link possibility as a result of the analysis. Stemming from the listed tasks, the last task of the work was generating a procedure and a new methodology supported by specifically designed tools (BIM-to-BRICK) to accomplish the aforementioned objective. Figure~\ref{fig:TaskWorkflow} represents a visual structure of the task sequence.

The paper is organized as follows: Section \ref{sec:methods} introduces the proposed methodology for the process and add-in creation in translation between BIM and its corresponding BRICK model. Section \ref{sec:application} presents its application and evaluation of a study building, and Section \ref{sec:discussion} performs a discussion of the main findings. Eventually, Section \ref{sec:conclusion} reports on the conclusions and remarks.

\section{Methodology}
\label{sec:methods}

This section outlines the main features of the methodology proposed to create a direct bi-directional relationship between a BIM model and an associated BRICK model. The chapter begins by giving an overview of the introduced workflow, then the add-in BIM-to-BRICK is described in further detail. 

\subsection{General description}
\begin{figure*}[h]
\begin{center}
\includegraphics[width=0.95\textwidth, trim= 0cm 0cm 0cm 0cm,clip]{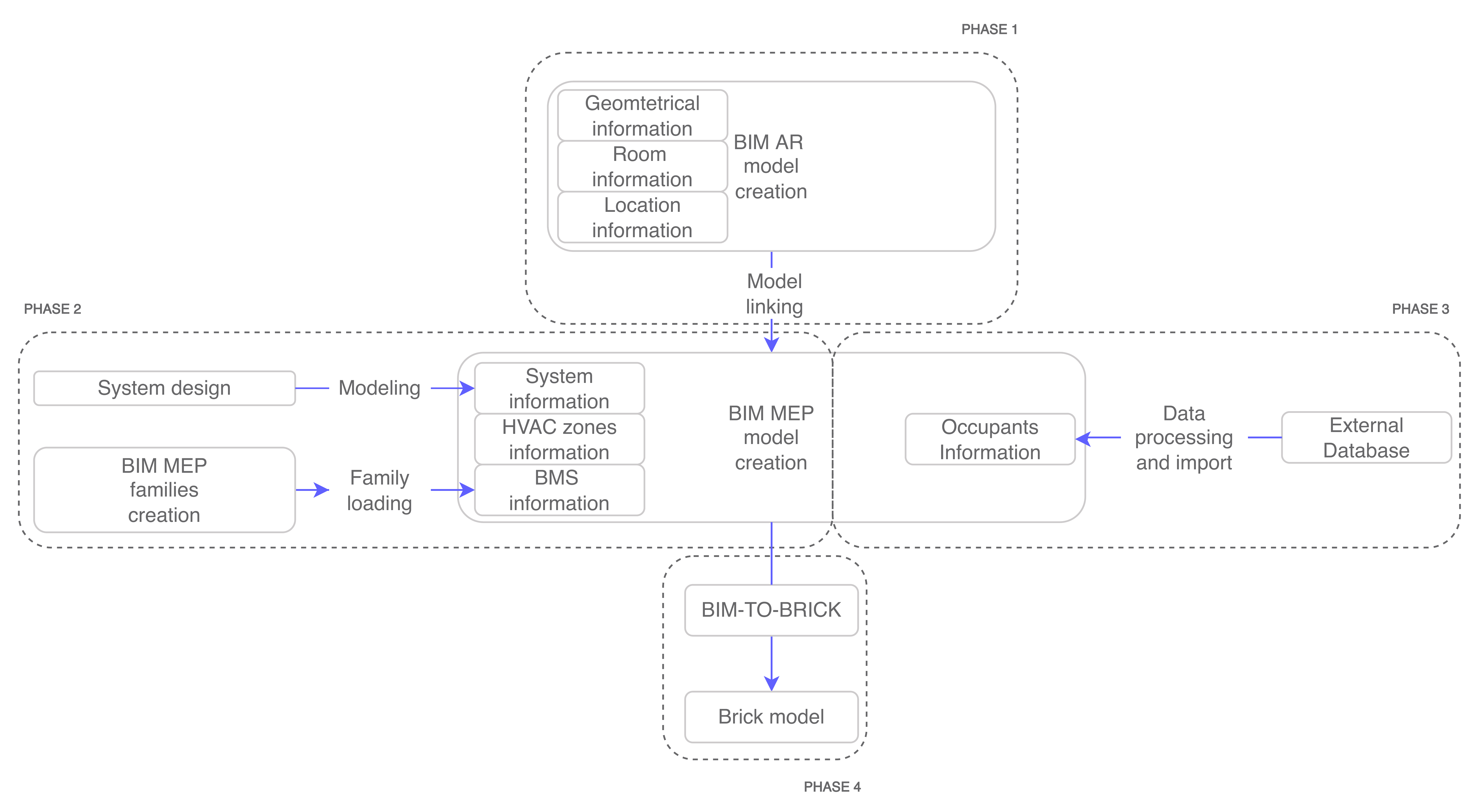}
\caption{Main phases of the workflow for BIM-to-BRICK model generation}
\label{fig:MethodWorkflow}
\end{center}
\end{figure*}

The main aim of the proposed work is to develop a methodology (and associated tools to make it applicable) focused on the automated process of convergence between the BRICK and BIM data structures -- a process that manually would take huge amounts of manual effort. The methodology is developed using Autodesk Revit, a commercial building information modeling (BIM) platform provided by Autodesk. Dynamo, a visual programming plug-in for Autodesk Revit, and its embedded CPython engine are used to generate the BIM-to-BRICK add-in, which converts a BIM model to a BRICK model directly.  The general workflow is exemplified by Figure~\ref{fig:MethodWorkflow}, which consists of four main phases. In Phase 1, the support BIM architectural model is created, then a second BIM MEP (Mechanical, Electrical Plumbing) model is generated in Phase 2. The architectural model defines the geometrical aspects of the building, identifying the main component such as floors, walls, windows, and rooms.
The BIM methodology allows every object within the project model to be enriched with information that characterizes the elements. The data can be related to geometrical or quantitative aspects and then be used to identify relationships between the components.

A link between the two entities is established to coordinate the development of the baseline for the digital twin. Different models allow users to focus specifically on the targeted discipline, enabling a more productive real-time collaboration and more organized information in independent environments. Complementing the architectural model, the MEP model includes system-related data such as ducts, VAV, fan coils, and thermostats into BIM. BMS-related information is inserted as part of the MEP model with custom Revit families developed explicitly for this work. The parameters of the custom families allow users to insert the relevant data such as an identifier (fundamental for establishing a one-to-one relationship between the BIM and BRICK framework), hosting room, time series identifier (useful to access external data), and master panel name. The created parameters aim at identifying the roles, associations, and information of the BMS points. In Phase 3, as part of the MEP model, occupants-related information in external datasets is added to the BIM environment. Such data can be obtained from wearable devices and thanks to human comfort surveys. Since the occupants do not have a physical representation within the BIM model, this external data is programmatically accessed using Dynamo. A single environment with the required information from architectural, MEP, and occupant-related data is established at this stage. In Phase 4, BIM-to-BRICK, a Revit add-in to convert the BIM model resulting from the first three phases into a BRICK model is created. A Revit identifier is included for all the entities in the delivered BRICK model, allowing users to generate, modify and update the result at any moment along the development of the digital twin.

\subsection{Add-in description}
BIM-to-BRICK is the add-in application that can be executed to conveniently convert a BIM model into a BRICK model, generating an optimized and user-friendly workflow. Compared to existing processes of referring and translating from the BMS point labels or creating a table of the different entities and relations, it automatically infers their metadata from its location and associations in BIM. To ensure the scalability of the provided tool, it is created using Revit API and built-in parameters. The Built-in parameters, or OOTB (Out of the box) parameters, are provided by default within a project or family file. Thus, they cannot be changed or deleted when this framework is employed. This feature is included to improve the robustness and resilience of the tool, which can thus be used from multi-language authoring platforms to exchange information consistently.

% \newpage
\subsubsection{Integrating BMS and systems information}
\label{sec:bmsinfo}

This methodology relies on default Revit components and custom Revit families to integrate BMS information inside the BIM environment. The Revit database is organized into four levels of hierarchy: category, family type, and instance. A Revit family is a set of elements in a specific category. It is not only a component of the project, but also the carrier of various additional parameters and information ~\cite{7825085}. The default BIM components of the architectural model provide basic information about the building that will be defined in the resulting BRICK model, such as the name of the building, its floor levels, and the general context for the construction site. In addition to the elements provided in the architectural model, ducts, air terminals, spaces, and associated HVAC zones are added in the MEP model. These default elements are essential to create the context for the BIM-to-BRICK add-in to understand the layout and distribution of the BMS system.  The core information about the BMS entities, such as element relationships, time series identifiers, and control logic, is then integrated with custom families. Within this work, three representative objects are generated to address the requirements of a basic BMS system: Variable Air Volume box (VAV), Fan Coil Unit (FCU), and Thermostat. The developed elements can be enriched with BMS-related information. Data connectors, a native tool of the used technology, allow users to define the link between the used components.  An example can be represented by a VAV box controlled by (and thus, linked to) a thermostat inside a generic room. In other words, connectors allow expression in terms of Revit API, the control logic between the BMS elements. VAV and FCU families exploit the rectangular duct connector to link to the duct net. All the families are equipped with room calculation points used within the Revit project environment to calculate the location of the single component inside the building. This information helps infer how the HVAC zones are related to the rooms and other BMS components. 
 
% \newpage
The developed elements can be enriched with BMS-related information. An example can be represented by a VAV box controlled by (and thus, linked to) a thermostat inside a generic room. In other words, connectors allow expression in terms of Revit API, the control logic between the BMS elements. VAV and FCU families exploit the rectangular duct connector to link to the duct net. All the families are equipped with room calculation points. Room calculation points are used within the Revit project environment to calculate the location of the single component inside the building. This information helps infer how the HVAC zones are related to the rooms and other BMS components. 
The custom families host custom project parameters within the project MEP environment. Project parameters are containers for information defined by the user and then added to multiple categories of elements in a project. These parameters are defined explicitly during this work to allow users to easily input information regarding the components and time series identifiers that can be embedded in the resulting BRICK model. Figure~\ref{fig:GeneralWKF} shows an overview of all the elements, properties, and relationships involved in the proposed methodology:

\begin{figure*}[h]
\begin{center}
\includegraphics[width=0.85\textwidth, trim= 0cm 0cm 0cm 0cm,clip]{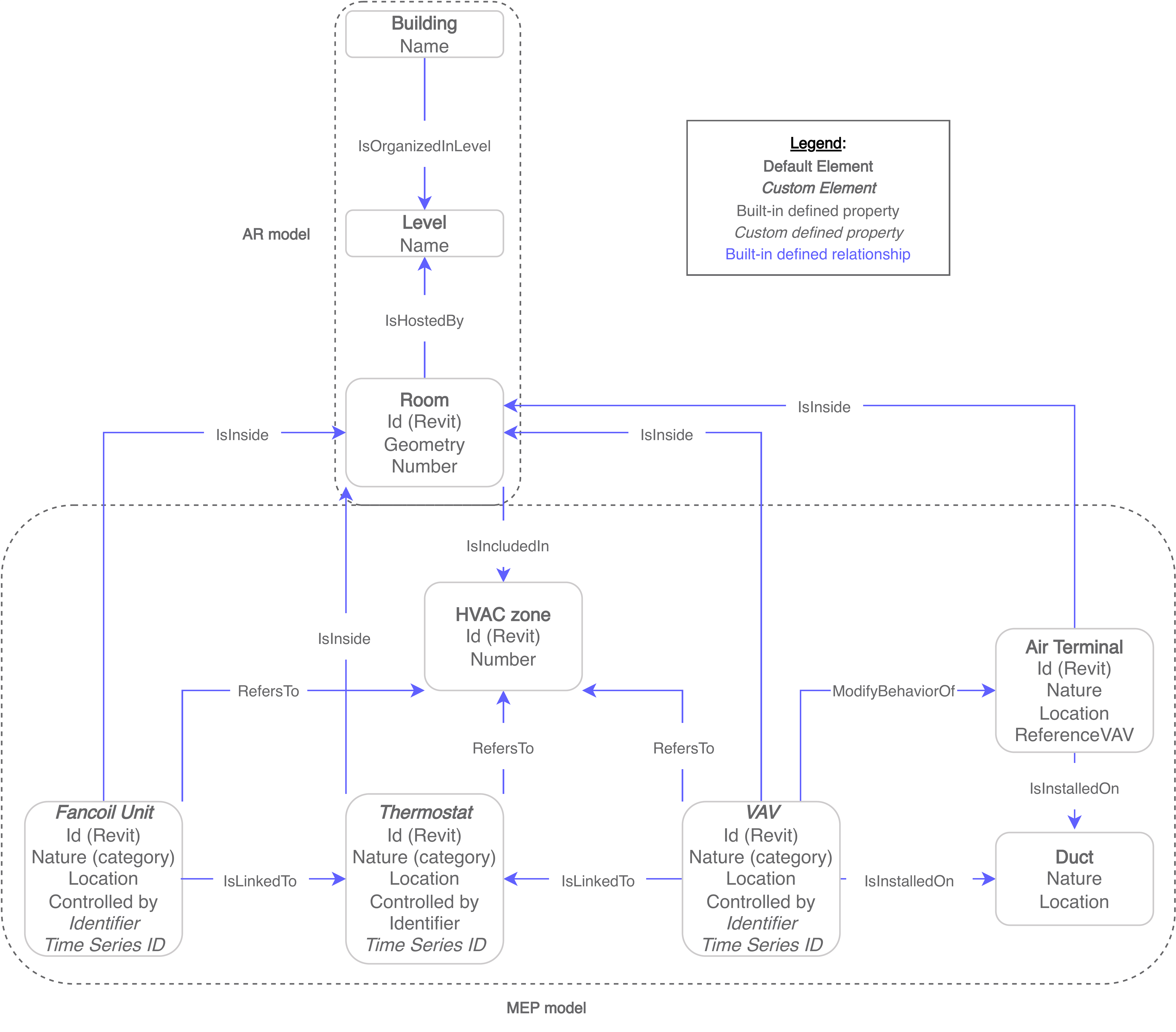}
\caption{Programmatically inference of elements, properties, and relationships of BMS entities in BIM}
\label{fig:GeneralWKF}
\end{center}
\end{figure*}

% \newpage
\subsubsection{Integrating occupants' information}
\label{sec:occinfo}

Integrating occupant data within a digital twin of a building invest a pivotal role and highlights the importance of this approach in enhancing building performance, occupant comfort, and overall sustainability. By considering occupant behavior, preferences, and needs, the digital twin can provide valuable insights for optimizing building design, operation, and energy management strategies. The integration of occupant data in the framework not only enables a more accurate representation of real-world conditions but also facilitates data-driven decision-making processes for building stakeholders. In the current methodology, occupants-related information is stored within an external dataset, and occupants do not have a physical representation inside the existing models. Intensive longitudinal methods have begun to emerge as a way to characterize occupants for various built environment objectives~\cite{Jayathissa2020-pv}. This section of work focuses on integrating intensive longitudinal data within the BIM-to-BRICK conversion workflow.
The procedure's first step builds upon the BRICK schema's extension to represent the metadata of occupants provided by Luo et al.~\cite{LUO2022104307}. The instances of the occupant data are programmatically generated from the external dataset to BIM. For each occupant added to the model, the procedure defines the general information of the subject according to the extension of the BRICK schema for occupants (i.e., age, gender, ethnicity…). The second step of the procedure focuses on finding the occupant's location inside the building. The provided methodology is intended to convert the Geographic Information System (GIS) automatically coordinates system (latitude-longitude) to the local coordinate system (X, Y, Z) to define the spatial representations of each entity directly within the Revit environment. A preliminary manual process is required to find the relative roto-translation and the scale difference between the actual construction and the modeled building. In addition, the source database is supposed to indicate the altitude of the measured point to locate the latitude and longitude coordinates in a three-dimensional space. When the underlined preliminary requirements are met, a python routine based on the UTM python package~\cite{bieniek2016bidirectional}, included within the BIM-to-BRICK add-in, links the collected latitude-longitude data to the Revit database. The occupants’ location is represented as a point object in the Dynamo environment. This solution allows the user to identify geometrical relationships between the occupants and the building components (i.e., rooms, HVAC zones…) without creating redundant BIM objects.

\subsubsection{BIM-to-BRICK software description and UI}

The methodology described in Section 2 is presented as a convenient Revit add-in called BIM-to-BRICK. BIM-to-BRICK can generate a bi-directional link between BIM and BRICK models. Figure~\ref{fig:DynamoCanvas} shows the developed code embedded within BIM-to-Brick:

\begin{figure*}[h]
\begin{center}
\includegraphics[width=0.9\textwidth, trim= 0cm 0cm 0cm 0cm,clip]{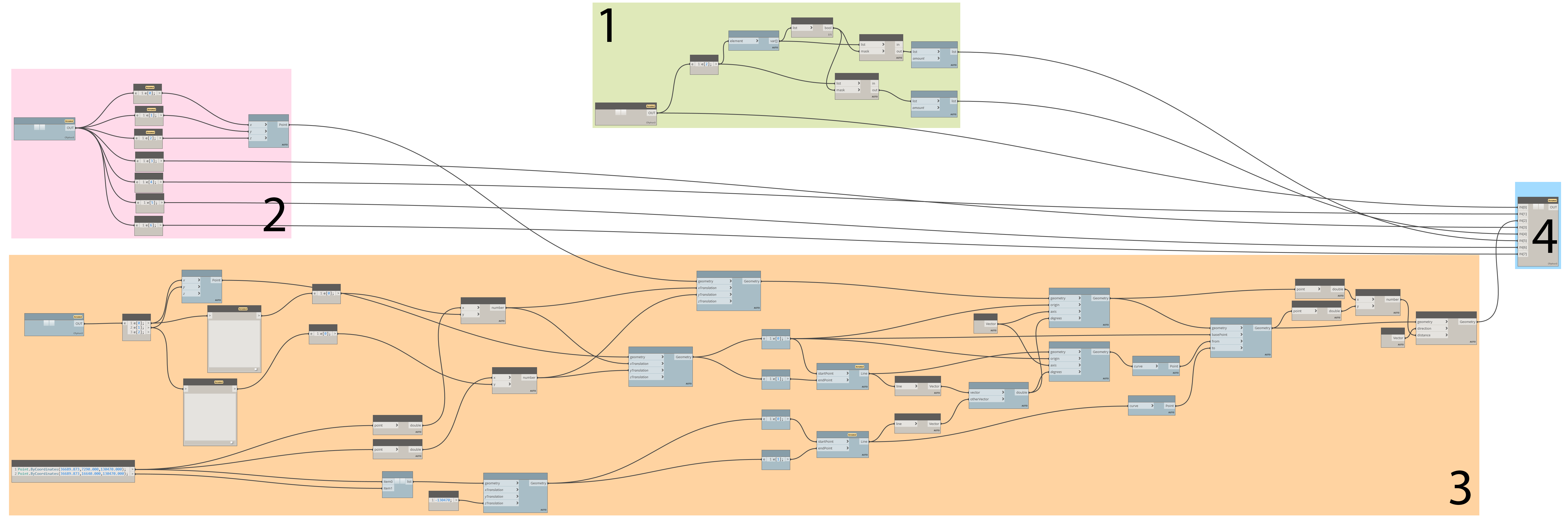}
\caption{High-level schematic of the Dynamo canvas showing the BIM-to-BRICK developed logic}
\label{fig:DynamoCanvas}
\end{center}
\end{figure*}

The add-in is developed inside Autodesk Dynamo by combing pre-set nodes with custom-developed Python nodes. The developed result can be discretized in four main groups of performed actions, as shown in Figure~\ref{fig:DynamoCanvas}:

\begin{enumerate}
    \item Data collection from an external database
    \item UTM conversion, translating latitude-longitude-altitude coordinates to (x,y,z) coordinates 
    \item Collection of data from the Revit database 
    \item Interpolation of data and final BRICK model generation.
\end{enumerate}

The generated link can be defined as bi-directional because every object existing in the BIM model and included in the conversion obtains a traceable BRICKified analog at the end of the process. Every element in a Revit model has a unique element identifier used to identify elements in the model. The created BRICK elements preserve the ID deriving from the BIM model, allowing the user to trace every single created BRICK object back in the source BIM model.

The main objectives of BIM-to-BRICK are:

\begin{itemize}
    \item Allow fast and efficient generation of BRICK models based on BIM models
    \item Allow users to edit and review the BRICK model of a building promptly
    \item Provide an intuitive tool to generate BRICK models
    \item Provide a new link between the BIM and BRICK paradigm
    \item Move a step towards the concept of digital twin
\end{itemize}

\begin{figure*}[h!]
\begin{center}
\includegraphics[width=0.9\textwidth, trim= 0cm 0cm 0cm 0cm,clip]{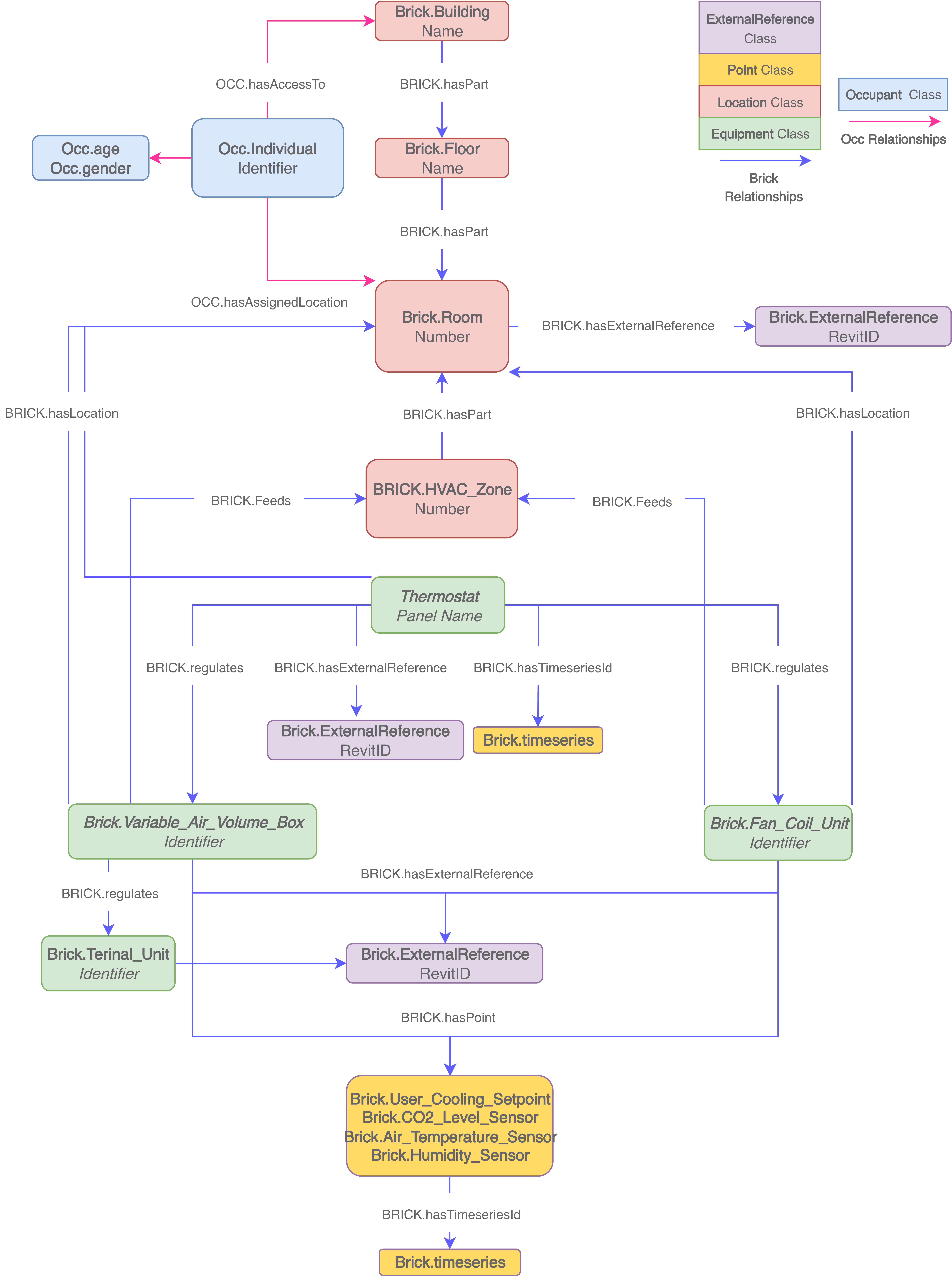}
\caption{Overview of the BRICK classes, instances, and parameters considered within BIM-to-Brick}
\label{fig:NewWF}
\end{center}
\end{figure*}

An integrated BRICK model associated with the modeled instances is generated using the add-in described in Sections \ref{sec:bmsinfo} and \ref{sec:occinfo}. No additional inputs are requested from the add-in to create the BRICK model. A more complex BRICK model of the building is obtained when all the custom parameters shown in Figure~\ref{fig:NewWF} (BIM relationships) are filled with the required information. 
Figure~\ref{fig:NewWF} shows an overview of the BRICK classes and instances (including the occupancy extension) obtained at the end of the conversion.

BIM-to-BRICK adds to the basic functionality of Autodesk Revit with a dedicated tab. The graphical user interface is divided into two main sections. Starting from the left, a preliminary set of user-defined parameters is provided. Then on the right, in the convert section, three push buttons allow users to generate the BRICK model. The \emph{BMS} push button can create a BRICK model involving all the building equipment and building management system (BMS). Similarly, the \emph{People} push button can create a model of the occupants involved within the given analysis. Lastly, the \emph{DigitalTwin} push button combines the previous ones, obtaining a BRICK model of the complex system made of building, occupants, and their interactions. Table I collects an overview of the possibilities offered by BIM-to-BRICK.

\begin{table*}[h!]
\begin{center}

\begin{tabular}{ |p{4cm}|p{12cm}|  }
 \hline
 \multicolumn{2}{|c|}{BIM-to-BRICK modules capabilities} \\
 \hline
 Module& Involved objects\\
 \hline
 BMS   &  Building equipment and Building Management System (BMS) of the targeted systems\\
 People&  Dataset of occupants involved within the analysis\\
 Digital Twin &Building equipment and Building Management System (BMS) of the targeted systems, dataset of occupants involved within the analysis\\
 \hline
\end{tabular}
\caption{ BIM-to-BRICK different modules overview}
\end{center}

\end{table*}
% $\\$

\begin{figure*}[h]
\begin{center}
\includegraphics[width=0.9\textwidth, trim= 0cm 0cm 0cm 0cm,clip]{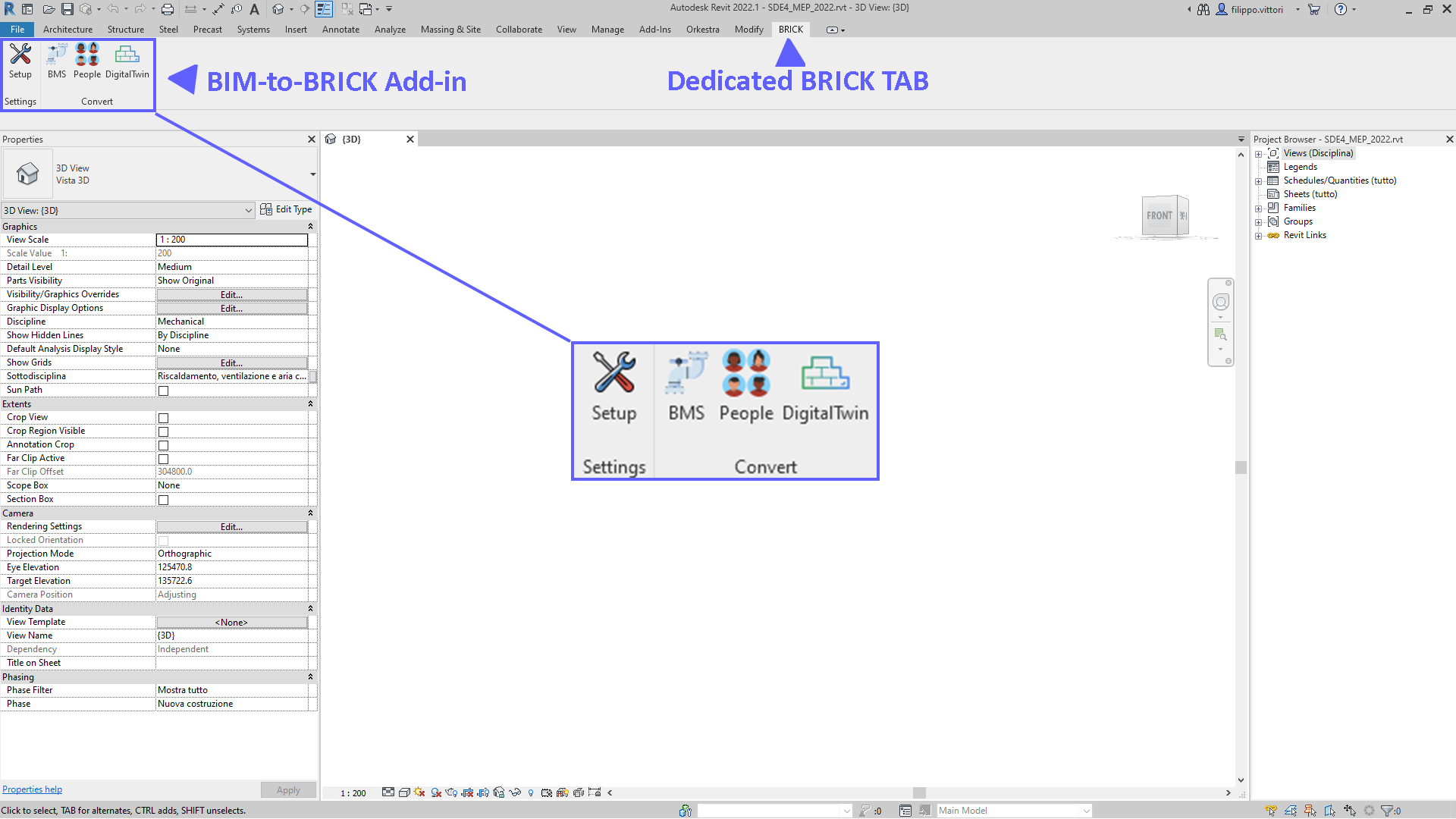}
\caption{BIM-to-BRICK graphical user interface}
\label{fig:Interface}
\end{center}
\end{figure*}

A Terse RDF Triple Language (Turtle) file of the BRICK model is generated and stored in a target folder by pressing one of the commands to execute the process. Simultaneously, the add-in automatically proposes to the user BrickStudio~\cite{brickstudio}, a web-based RDF visualizer able to parse and generate a visual representation of any remote RDF resource. Eventually, the turtle file can be loaded in BrickStudio (or other BRICK rendering platforms) to obtain a graphical representation. 
Figure~\ref{fig:Interface} represents the developed graphical user interface.

\section{Implementation and Results}
\label{sec:application}

This section describes the application of the provided methodology and the implementation of BIM-to-Brick. The chapter begins by identifying and elaborating on the main features of the building and database used. Finally, the obtained results and potential future applications are identified.

\subsection{Case study: SDE4 – The building and the systems}
SDE4 is a building located in Singapore within the NUS School of Design and Environment (SDE). This six-story building has an overall floor area of 8588 m², hosting several functions, including providing office spaces, lecture halls, laboratories, and co-working environments. It consists of 86 zones, many of which are naturally ventilated (NV), including transitional (e.g., corridors), semi-open spaces (e.g., the plaza and terraces), and service rooms (e.g., toilets and storage). Other zones are either mechanically ventilated (MV) or use traditional air-conditioning (AC). Many spaces adopt a hybrid cooling system (HC) that employs both AC and ceiling fans with a design zone set point temperature of 27 degrees Celsius and elevated air movement via ceiling fans~\cite{abdelrahman2022personal}. The building’s systems are linked and implemented through a BMS to control and manage the system operations. SDE4 is a prototype of net-zero-energy and sustainable design thinking in the tropics. The building has been awarded a Green Mark Platinum certificate~\cite{zhou2019challenges}.

\subsection{Database: Intensive Longitudinal Indoor Comfort dataset - People}
\begin{figure*}[h]
\begin{center}
\includegraphics[width=0.9\textwidth, trim= 0cm 0cm 0cm 0cm,clip]{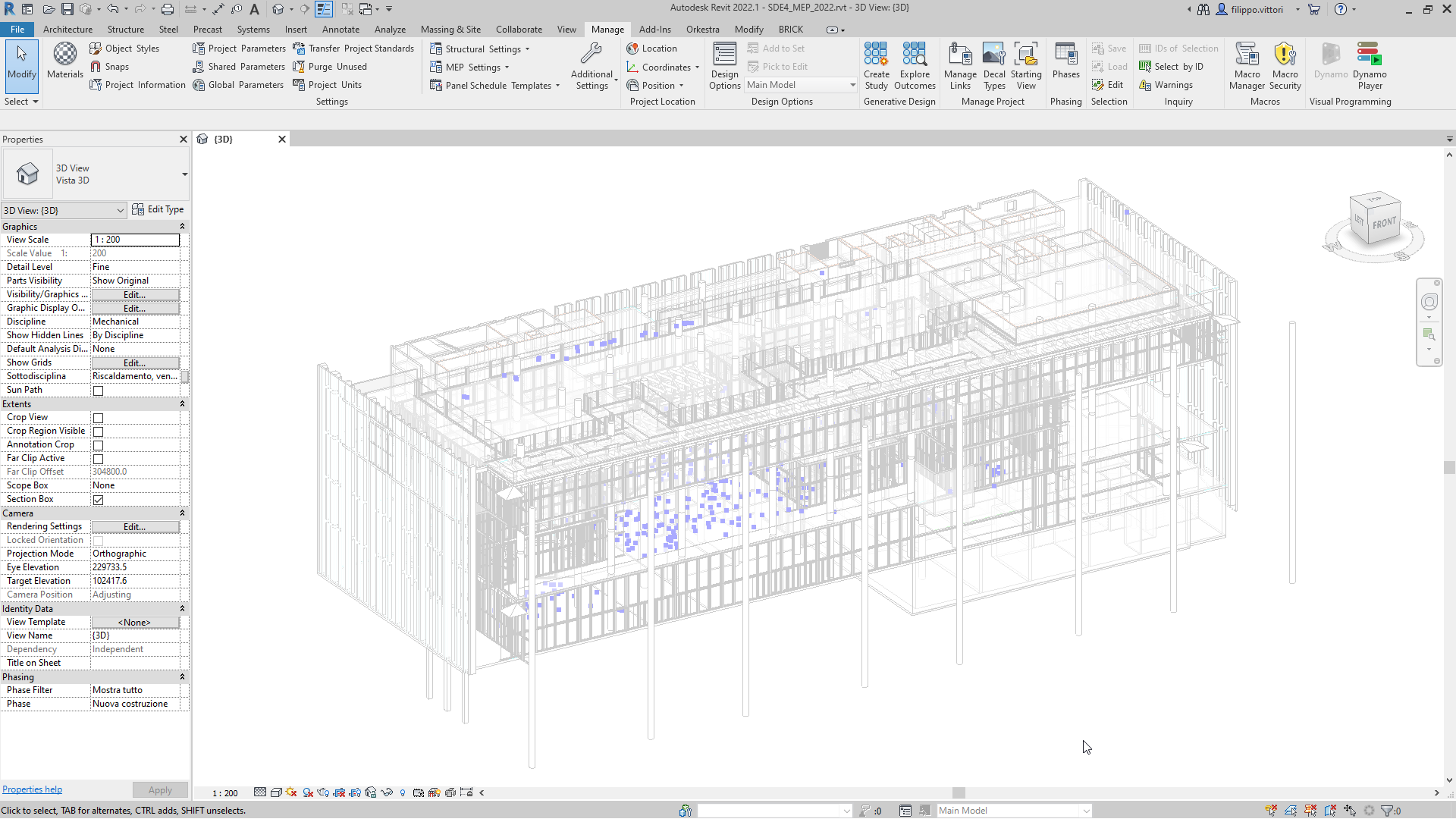}
\caption{Occupant locations reflected as representative points objects within the BIM model}
\label{fig:ModelPoints}
\end{center}
\end{figure*}

This work stems from a longitudinal dataset from a previous study involving 30 participants. The subjects involved in the experiment expressed using the Cozie application~\cite{Jayathissa2019, Tartarini2022-iu} dynamic preferences about their thermal, lighting, and noise-based perception~\cite{Jayathissa2020-pv}. The study also collected temporal data from fixed IEQ sensors in the indoor environment (temperature, humidity, noise, and carbon dioxide) and physiological data from the smartwatch (heart rate, near body temperature). This study is one of several similar recent efforts to collect longitudinal subjective occupant data in the built environment using smartwatches~\cite{Miller2022-sj, Miller2022-dy, Quintana2021-ka} The delivered software uses the information retrieved from the database to characterize the experiment participants in terms of identifier, age, and gender. The BRICK classes used are the \emph{Occupant} class and the relative \emph{individual} subclass. Class properties are used to define the age and gender of the individuals ~\cite{LUO2022104307}. To avoid non-defined values deriving from the experimental campaign, the experiment focuses on 17 participants. Occupants’ location is collected through a couple of latitude and longitude coordinates via Bluetooth Low Energy (BLE) beacons. Further descriptions of the experiment settings and data collection are provided by Abdelrahman et al.~\cite{abdelrahman2022personal}. The latitude and longitude measurements are coupled with every measurement within the cited experimental study to give the coordinates to be converted and locate people inside SDE4, according to the conversion procedure described in Section \ref{sec:occinfo}. Figure~\ref{fig:ModelPoints} illustrates the result of the processed locations of occupant information in Revit. Every blue point represents a subject’s location at a given timestamp.

% \newpage

\subsection{BIM-to-BRICK: Application on SDE4 building}
SDE4 building is used to demonstrate the proposed methodology to assimilate its data with the longitudinal personal comfort dataset. The automatically generated BRICK model harnesses the information input within the linked BIM model of the building, consisting of the architectural and MEP discipline. The file size of the architectural model is 30,760 KB, while the file size of the MEP model of the building is 15,784 KB. The geometrical aspects and the location of the objects inside the building are accounted for in the framework of the architectural model. In contrast, the MEP model provides information about the mechanical systems and related BMS controls. The longitudinal database completes the framework with the required data associated with the human side of the delivered result, leveraging the cited extension of the BRICK schema to represent metadata of occupants.
The combination of the SDE4 BIM model and the selected database allows BIM-to-Brick to generate the RDF-based model. The add-in equips the user with the option to create three different models depending on the information included in the conversion. Within this first application of the procedure, all three options are generated and investigated in the graph model visualizer. All the entities and logical relationships involved in both the dataset and the models are converted to the new RDF-based form, following the schema reported in Figure~\ref{fig:NewWF}.
The different information embedded within the models generates different shapes and complexity of the networks. The “People” version of the model shows a relatively small and straightforward chart involving only the characterization of the subjects and the relationship with the experimental environment generated within the RDF schema. This version embeds one building, nine building levels, 48 rooms, and HVAC zones, and 17 individuals. The generated code filters only the relevant subjects to be considered within the analysis (as underlined in Section \ref{sec:occinfo}), obtaining the final considered sample of 17 subjects. The resulting BRICK model is generated in 11 seconds. 

\begin{figure*}[h]
\begin{center}
\includegraphics[width=0.85\textwidth, trim= 0cm 0cm 0cm 0cm,clip]{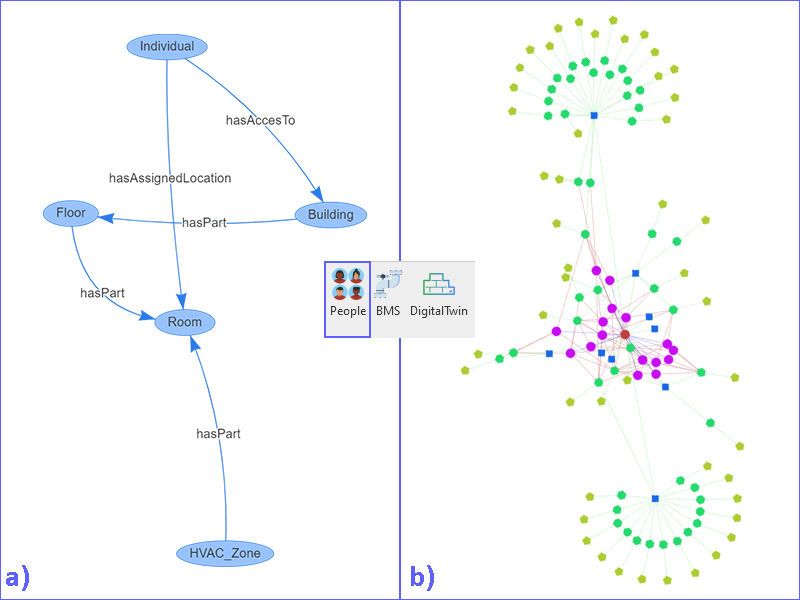}
\caption{A \emph{People} BRICK model generated by BIM-to-BRICK: a) A synthetic view of the BRICK classes involved is automatically generated by the BRICK Viewer, and b) Overall view of the rendered BRICK model within BrickStudio}
\label{fig:People}
\end{center}
\end{figure*}

\begin{figure*}[!htbp]
\begin{center}
\includegraphics[width=0.85\textwidth, trim= 0cm 0cm 0cm 0cm,clip]{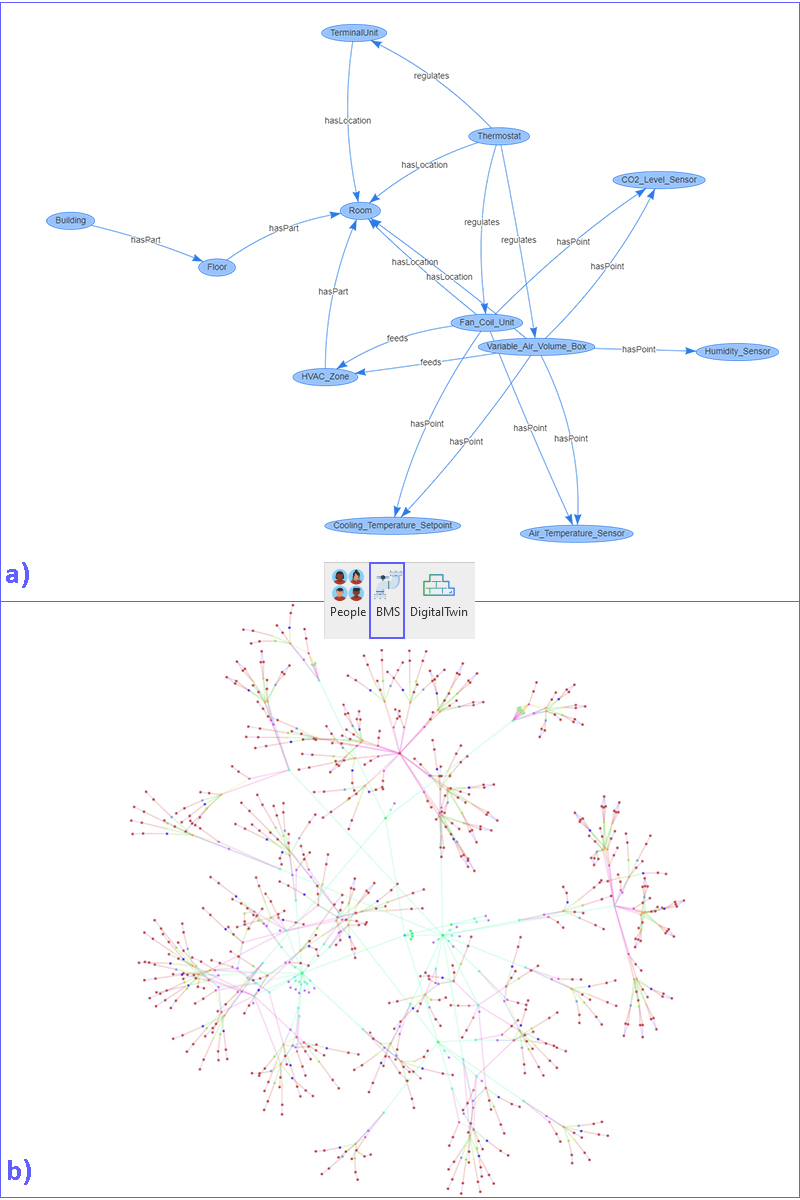}
\caption{A \emph{BMS} BRICK model generated by BIM-to-BRICK: a) A synthetic view of the BRICK classes involved is automatically generated by BRICK Viewer, and b) Overall view of the rendered BRICK model within BrickStudio}
\label{fig:BMS}
\end{center}
\end{figure*}

\begin{figure*}[!htbp]
\begin{center}
\includegraphics[width=0.85\textwidth, trim= 0cm 0cm 0cm 0cm,clip]{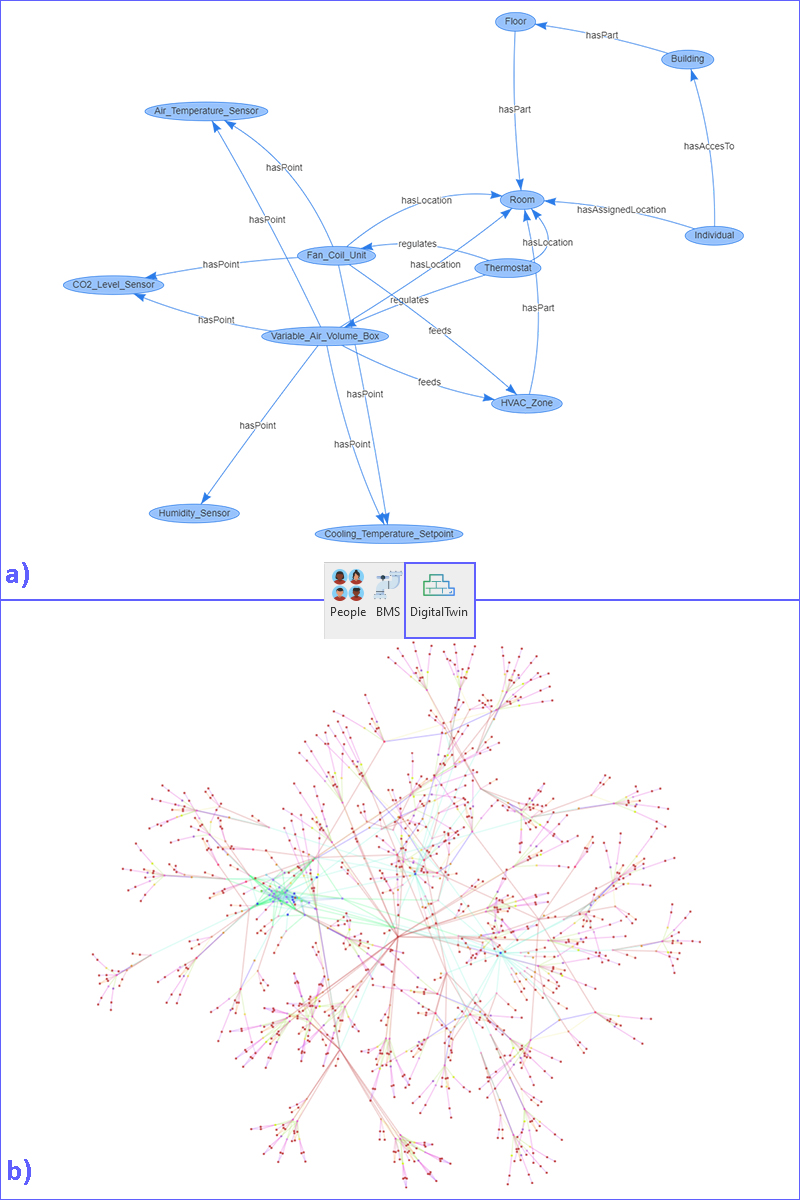}
\caption{A \emph{Digital Twin} BRICK model generated by BIM-to-BRICK: a) A synthetic view of the BRICK classes involved is automatically generated by BRICK Viewer, and b) Overall view of the rendered BRICK model within BrickStudio}
\label{fig:DT}
\end{center}
\end{figure*}

Figure~\ref{fig:People} shows the classes and relationships created by BIM-to-BRICK and an overall view of the rendered BRICK model.
The “BMS” version of the model shows a more complex network involving the characterization of the mechanical and related BMS system, also capturing the relationships between the two entities. The information retrieved from the architectural BIM model is crossed with the data deriving from the MEP discipline, allowing the systems to relate to the building's physical aspects. This version embeds one building, nine building levels, 48 rooms, and HVAC zones, two fan coil units equipped with CO2 sensors and thermometers, 14 fan coil units equipped with thermometers, and 52 VAV boxes equipped with CO2 sensors, hygrometers, and thermometers. The resulting BRICK model is generated in 15 seconds.  Figure~\ref{fig:BMS} shows the classes and relationships created by BIM-to-BRICK and an overall view of the rendered BRICK model. The \emph{Digital Twin} model shows the complete version of the BRICK model that BIM-to-BRICK can provide. This solution involves all the components deriving from the BIM models and the dataset, creating all the relationships described in Figure~\ref{fig:NewWF}. This version embeds one building, nine building levels, 48 rooms, and HVAC zones, two fan coil units equipped with CO2 sensors and thermometers, 14 fan coil units equipped with thermometers, and 52 VAV boxes equipped with CO2 sensors, hygrometers and thermometers, and 17 subjects. The resulting BRICK model is generated in 17 seconds. Figure~\ref{fig:DT} shows the classes and relationships created by BIM-to-BRICK and an overall view of the rendered BRICK model.

% \vfill
% \vfill

% \newpage

% \newpage

% \newpage

\subsection{Potential applications of the SDE4 BIM-to-BRICK framework}
The first application of the delivered methodology provides a BRICK model of the considered parts of the SDE4 building, starting from the existing BIM models. This sub-section underlines potential applications and related outcomes of the obtained result.

\subsubsection{Empowering the information of the models}
The delivered methodology can provide a BRICK model starting from an existing BIM model.  BIM models are commonly offered within the design process of a building because they are the environment in which designers of various disciplines collaborate and create the final drawings needed for the construction phase ~\cite{Ashcraft2008}. Creating a BRICK model from scratch is possible thanks to understanding the project documents and the manual translation of the information within the RDF-based language. With the provided methodology, all the information contained within the BIM models is automatically converted, obtaining the same result as what is accomplished with a manual process. The logical relationships between the component of the BMS system can easily be input by designers within the last phase of the design and construction process, mitigating the need for highly specialized human capital wholly dedicated to the task. Once the information is correctly set within the BIM model, the conversion can be performed in seconds instead of days or weeks.

\subsubsection{Instant updating of the BRICK model and visual interaction}
The created framework allows users to interactively link the information present in the BIM and the BRICK model. A BRICK model usually presents itself in two possible formats: a series of lines of code or a visual representation (graph format). The BIM model equips users with 3D environments where components can be easily visualized, located, and edited. Because of the provided bi-directional link of the information, the object within the BIM models presents a unique analog within the created BRICK model. The identifier parameter plays a fundamental role, allowing users to focus on the same element in both BIM and BRICK models. Consequently, both models can be modified and updated smoothly. Figure~\ref{fig:ID} exemplifies the concept of the ID parameter for a thermostat instance:

\begin{figure*}[h]
\begin{center}
\includegraphics[width=0.7\textwidth, trim= 0cm 0cm 0cm 0cm,clip]{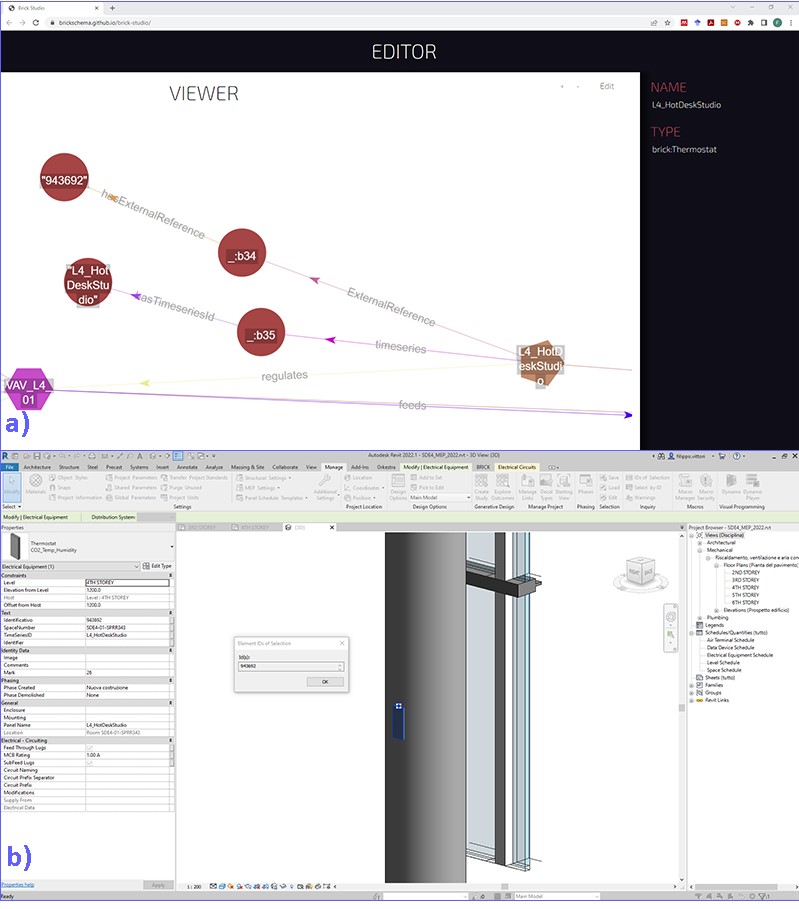}
\caption{The ID parameter as the link of the two platforms: a) ID parameter within the rendered BRICK model inside BrickStudio, and b) ID parameter within the source BIM model}
\label{fig:ID}
\end{center}
\end{figure*}

% \newpage

\subsubsection{Standardizing the approach to extend the framework}
The delivered methodology sees the first application to the SDE4 building, empowering the existing BIM model. Still, the workflow and the related add-in can be applied to every building information model set according to the information shown in Section 2. The SDE4 can be the first case-study model to use as a template for future implementation in other buildings.

\section{Discussion}
\label{sec:discussion}
The proposed work opens new possible developments related to the integration of the data deriving from the buildings’ design, construction, and operation phases. This section discusses outcomes and the potential value for researchers and professionals deriving from the workflow application and related BIM-to-BRICK add-in. At the end of the chapter, limitations of the current state of the framework are underlined and discussed, posing the basis for future developments of the work.

\subsection{Bridging data silos with common languages}
The main aim of the current work is to define a linking workflow between BIM models and BRICK schema. This effort moves towards avoiding silos of information created by the difficulty of exchanging information generated by different sources but with a common denominator: the considered object. Data interoperability is thus a key component, and using semantic web technologies such as resource description frameworks is a suitable approach. Interoperability between different platforms and formats represents the current main challenge to exploiting the vast amount of information that the AEC (Architecture, Engineering, and Construction) industry is generating ~\cite{Tang2020, Ren2021, Luo2021}.

\subsection{Automation of data model integration}
This study leverages the capability of BIM to manage related data with automation routines. The paradigm introduced with BIM changes the designers’ way of managing and authoring the design. Building a model of information is not drawing. The drawing document is one piece of information that can be extracted from the model. This approach opens new possibilities to generate some parts of the design automatically. Several studies focus on the possibility of automating repetitive tasks, reducing the time and resources needed across all the design, construction, and operation processes ~\cite{Zhang2022}. The control logic can be embedded by creating an automatic process, generating workflows that augment the base capability of the authoring software used to design the future building. New rules and features can be added ~\cite{Wijayakumar2013, Abrishami2020}, and new sources of information can be embedded within the resulting output~\cite{Olsson2018, Banfi2017}.

% \subsection{Digital twin, BIM, and BRICK}
The digital twin concept is gaining momentum and rising interest as one of the multiple forms of the current digital transformation ~\cite{Wagner2019}. The framework outlined is inserted within the definition of a digital twin as a possible answer to the research and industry’s demands to closely monitor and control assets, providing real-time insights into a building’s performance and events. The proposed approach wants to underline the importance of the capability to convert information from one form to another, moving toward an inclusive, dynamic environment. The integration between BIM models and other languages, such as a graph, RDF, or GNNs, represents a further step in the digital twin direction~\cite{GNECCO2023112652, Xie2022}.

\subsection{Limitations of the methodology}
The methodology and tool provided within the current work primarily demonstrate how it is possible to carry out an efficient and rapid conversion, establishing a two-way link between BIM and BRICK models. 
The current version of the add-in is developed inside a proprietary platform (Autodesk Revit), preventing at the moment direct usage inside other vendor's platforms (Bentley, Nemetschek, Trimble, etc.). The capability of Autodesk Revit to manage and export IFC (Industry Foundation Class) models, including the relevant information for the BRICK conversion, is the base reason for the software choice to allow the development of the work in an interoperable framework. The methodology can still be applied, but the add-in BIM-to-BRICK should be replaced with alternative workflows.
The considered classes and BIM objects are limited to some examples that include critical features useful to show the potential of the delivered workflow and add-in. Many other objects and classes must be involved in the analysis to have a fully operable BRICK model. Another limitation of the workflow is the usage of custom parameters that let the users enter the required information to perform the conversion. The code is break-down into standardized modules related to BIM, BRICK, and external components, facilitating future developments of the code.

\subsection{Future developments}
Future developments will move further steps toward the interoperability of the delivered framework. BIM-to-BRICK is developed inside Autodesk Revit to allow the user’s authoring of the BIM model to be converted. This is the first step needed to equip professionals and researchers with usable tools. Within the interoperability framework, the next steps of the work will be focused on two main areas. The first area is related to refining the delivered solution, investigating the capability of the considered parameters to be quickly included within the IFC export process in a standardized way. The second area of study will further develop the Python-based created code with the final aim of creating a parallel version of the add-in independent from the Revit environment. This last deliverable will be focused entirely on the IFC format, aiming at converting models from every source BIM platform. 
The capability of the methodology to involve more complex systems is another field of needed development. The interaction between BIM models can cover more than a single building. The linking capability of the models allows users to create complex systems of related buildings in terms of location, geometry, and information ~\cite{Porwal2013}. The approach provided within this work can be enriched, including the conversion and the managing of the information deriving from a complex system of buildings. Furthermore, other data sources can be easily embedded within the provided routines: one example is the BRICK extension of the BRICK schema to represent metadata of occupants that, within this work, allowed the inclusion of the subjects’ experimental data residing on an external database. The provided methodology is conceived to be the first step toward expanding the framework to more and more complex systems, following the lead of the BRICK schema. In order to enhance the robustness and applicability of the delivered methodology, we acknowledge the significance of incorporating statistical analyses on future target samples involving real-world buildings and experiments. These analyses can provide valuable insights and validate the effectiveness of the proposed approach in real-world scenarios. In summary, future developments will be directed towards creating environments that prioritize interoperability, with a strong emphasis on incorporating the Industry Foundation Class (IFC) format into the process and exploring the possibility of utilizing the cloud as a direct host for both the model components and the final BRICK output. The cloud-based approach opens up opportunities for enhanced processing capabilities and further advancements in the field.

\section{Conclusion}
\label{sec:conclusion}
Data interoperability is an essential enabler for building operations’ technology to scale and adapt by removing the reliance on customized integration. Adopting a standard in both the buildings’ metadata definitions and BIM can facilitate the deployment of more efficient operations of its various lifecycle processes. However, current adoption is often tedious and requires some degree of manual intervention and interpretation.  In recognizing that this is one of the critical bottlenecks in the push towards a unified protocol, this project aims to provide a bypass to programmatically infer and extract all available entities in the existing BIM of a building to formulate its BRICK schema representation. The proposed BIM-to-BRICK work uses Dynamo, a visual programming plugin in Revit, to access all entities and their associations based on their location and system information within BIM. Upon this, an inference strategy is followed to determine the various BRICK entities and relationships required in building the BRICK model of the building. This approach is successfully demonstrated on SDE4 with the objects and associations of all available architectural, mechanical, BMS, and occupant comfort information mapped to the BRICK model. It is designed to be a straightforward and user-friendly tool for adopting such schema that can be used in deploying various portable building applications. Extending this approach to other schemas and standards should thus be possible by simply changing the inference strategy in this approach. Increasing the accessibility for onboarding an interoperable standard can pave the way toward more efficient building management, leading to a more adaptable and sustainable infrastructure outlook.
% \pagebreak
% \newpage
\bibliographystyle{model1-num-names}
\bibliography{references}

\end{document}